# Title: A value-focused thinking approach to measure community resilience

- Rohit Suresh Rohit.suresh1993@gmail.com

- Parastoo Akbari parastoo@iastate.edu
  Affiliation: PhD Student at Iowa State University

- Cameron A MacKenzie camacken@iastate.edu
  Affiliation: Associate Professor at Iowa State University

**Abstract**

Community resilience refers to the ability to prepare for, absorb, recover from, and adapt to disruptive events, but specific definitions and measures for resilience can vary widely from researcher to researcher or from discipline to discipline. Community resilience is often measured using a set of indicators based on census, socioeconomic, and community organizational data, but these metrics and measures for community resilience provide little guidance for policymakers to determine how best to increase their community's resilience. This article proposes to measure community resilience based on value-focused thinking. We propose an objectives hierarchy that begins with a community decision makers' fundamental objective for resilience. Six high-level objectives for community resilience—social resilience, economic resilience, infrastructure resilience, environmental resilience, availability of resources, and functionality of critical services—are broken down into measurable attributes that focus on specific outcomes that a decision maker would like to achieve if a disruption occurs. This new way of assessing resilience is applied to measure the resilience of an illustrative community to an improvised explosive device, a cyberattack, a tornado, a flood, and a winter storm.

**1. INTRODUCTION**

Several major disasters have occurred since 2010 such as Hurricanes Ian, Ida, and Laura, severe wildfires in the western United States, an historic cold winter storm in Texas, and the August 2020 derecho in the central United States. The frequency and the cost of these major disasters are increasing [1,2]. Disasters are uncertain, and it may be impossible to identify and prepare for every possible disaster scenario. Even if communities have prepared for specific disaster scenarios, each disaster behaves differently and can lead to deadly consequences and large financial costs, among other serious consequences. Increasing the resilience of communities can help these communities and their residents, neighborhoods, infrastructure systems, economies, and government services withstand and recover from disruptive events [3-5].

Community resilience [5-7], or disaster resilience, is commonly defined as "the ability to prepare and plan for, absorb, recover from, and more successfully adapt to adverse events" [8]. At the state, county, and community level, resilience is increasingly becoming part of emergency preparedness planning [9, 10]. Resilience measures are used to understand the risks of disasters in addition to, or perhaps instead of, more traditional risk metrics such as threat, vulnerability, and consequences [114, 115]. Measuring and quantifying community resilience is important for local policymakers because they need to determine where to allocate resources for emergency preparedness and how best to plan for emergencies so that their communities will be more resilient to disasters. Good methods to assess community resilience should help the emergency planning decision-making process.

Perhaps the most common way to assess community resilience is by selecting dozens of indicators that are typically categorized into several dimensions [6, 9, 11-15]. These indicators may be aggregated into a single number, a community resilience index. A community resilience index can be compared among different communities or counties to understand which geographic areas are more or less resilient to disasters. These indicators often are metrics for which data can be collected relatively easy (e.g., census data) although some resilience indices require surveys of residents.

A common feature of virtually all of these indicators or metrics for community resilience is that these indicators are inputs or characteristics of a community. For example, metrics might describe the socioeconomic status of community residents, the degree of home ownership, the number of civic or religious organizations within a community, or the size and revenue of businesses. Some studies have investigated if these indicators and indices are correlated with outcomes from a disaster such as property damage and fatalities [9, 16-18]. Indicators seem to be selected in part because they are relatively easy to measure, and data are often publicly available for these indicators. However, the availability of data should not be confused with the usefulness of that data for accurately describing resilience.

Community resilience indicators fail to provide guidance for how community leaders should make decisions regarding preparing for and planning for emergencies or how to increase their community resilience [5]. For example, if one of the indicators to assess community resilience is the percentage of residents with a high school diploma [13], should government officials attempt to increase the number of high school graduates in order to make their communities more resilient to disasters? Although increasing the number of high school graduates will likely have positive benefits for the community, it is not at all clear that such a strategy should be part of emergency preparedness and planning. These indicators may not provide meaningful ways for how to reduce the risk of a community to disasters [5, 116, 117].

This article proposes a new approach to assess community resilience that links community resilience measurement to decision making [5] and that identifies what community leaders and residents ultimately care about when an adverse event occurs. A value-focused thinking (VFT) approach [19] to measure community resilience is appropriate because of the multi-dimensional nature of community resilience [5]. Applying VFT to assess community resilience requires community leaders to identify fundamental objectives or outcomes that they want to achieve if a disaster occurs. Based on extensive research into the disaster and resilience literature and through conversations with some government officials, we propose dozens of outcome focused objectives. Several of these outcomes are the standardized impacts defined by the Federal Emergency Management Agency (FEMA) [118] to measure the impacts from threats and hazards. Focusing on outcomes helps to ensure that the metrics for community resilience align with what defines resilience. The VFT approach to community resilience can help policymakers understand the resilience of their community to different disruptive events. It allows them to compare the benefits of different alternatives for emergency preparedness. The approach outlined in this article provides more meaningful decision support for allocating resources to enhance resilience.

The rest of the article is as follows. Section 2 reviews previous research efforts that present various indices and methods to measure community resilience. Section 3 presents our method to measure community resilience, which applies the principles of value focused thinking. Section 4 illustrates the application of VFT by assessing the resilience of a hypothetical community to five hazards: an improvised explosive device (IED) attack, a cyberattack, a tornado, a flood, and a winter storm. Section 5 presents the conclusion of this article.

## 2. REVIEW OF LITERATURE

Resilience and social vulnerability indices can be used to help compare the vulnerability and resilience of different communities to disasters [6, 11, 12, 14, 20]. Resilience can be categorized into community capacities (social capacity, community functions, and planning) or assets (economic, social, environmental, and infrastructure) [13] and/or assessed along several dimensions: social resilience, economic resilience, institutional resilience, infrastructure resilience, and community capital [12,15]. The Community Disaster Resilience Index (CDRI) [9] uses a community's inherent capacities, called capitals, to define and measure resilience based on 75 different indicators categorized into social capital, economic capital, physical capital, and human capital. The Resilience Capacity Index (RCI) measures the resilience capacity, or the pre-disaster resilience level as an indicator of the potential performance of a location under stress [21]. Other examples include an economic resilience index [22] and the Conjoint Community Resilience Assessment Measure based on five factors: leadership, preparedness, collective efficacy, trust, and attachment to the place [23]. An framework for assessing community sustainable resilience is comprised three capital systems: social, economic, and environmental [119].

Resilience indices or community resilience assessments typically identify several metrics or indicators within each dimension, capacity, or function. Metrics are usually selected because they describe a community's preparedness or the inherent capacity of society or residents that should help the community withstand or recover from a disruption. For example, the percentage of residents who speak English measures the social resilience of a community; the ratio of large to small businesses describes economic resilience; and participating in hazard reduction programs is a metric for institutional capacity [6, 12]. The indicators are frequently normalized and assigned weights in order to aggregate them into a single number, a single measure of resilience. FEMA publishes the Resilience Analysis and Planning Tool that aggregates 22 indicators (e.g., unemployed labor force, percent of inactive voters, number of hospitals) that frequently appear in the resilience literature [120].

Many resilience and vulnerabilities indices are constructed and measured at the county level and populated by census data at the county level. Other approaches to community resilience include a city resilience framework [10, 24], developing a list of questions for a community to perform a self-assessment of its resilience [25], and using digital traces of population movement within a community during a natural disaster [26] Melendez et al. [27] suggest simulations based on game theory and machine learning and artificial intelligence could be integrated into community resilience frameworks to enable better decision making for disasters.

Despite the proliferation of resilience measurement framework and tools, no single tool fits all communities, which leaves local decision makers unsure of which, if any, of the community resilience frameworks to use [5]. Methods for selecting indicators, collecting data for these indicators, and weighting and aggregating these indicators into an index number may be so fraught with errors and uncertainties that policymakers should be very leery of using these indices for making decisions and allocating resources [28]. Community resilience should be framed within a more thorough understanding of the subjective views of the policymakers themselves, their local knowledge and culture, and the historical context of the place [7,29]. Socioeconomic indicators rely on census data that may quickly become outdated [30]. Aggregating indicators that measure completely different things may not be appropriate and may average or hide important extremes within these indicators [31]. In general, resilience measurement can be challenging because three dimensions should be taken into account: the temporal scale, the spatial scale, and the community capacities [32].

Some research has studied whether these measures of resilience correlate with the actual consequences a community experiences during and after a disaster [18]. The results of validation studies suggest these resilience assessments are generally negatively correlated with negative outcomes from disasters (e.g., deaths, property damage), but the results are also very mixed with significant variability. Aggregated scores for typical resilience dimensions were found to be statistically significant in predicting the recovery of communities after Hurricane Katrina, but the effects were very small [16]. More resilient Gulf Coast communities as measured by the CDRI experience less property damage and have fewer fatalities from floods but are also more likely to have more floods leading to fatalities [9]. Bakkensen et al. [17] find that two out of three resilience indices and two other vulnerability indices have predictive power for two of three outcome metrics (property damage, fatalities, and frequency of disaster declarations), but none have predictive power for all three outcomes. Statistical tests have also been performed for social vulnerability indices [33] and to quantify the relationship between social indicators and economic costs of disasters [34]. Ma et al. [35] find that higher scores for community resilience in China's earthquake-stricken areas are positively correlated with residents' overall emergency preparedness.

Another approach to assessing and measuring resilience appears in the engineering, infrastructure, and business literature [14, 36-38]. The foundational concept for many of these resilience assessments is the "resilience triangle" which measures resilience according to a sudden decrease in performance due to a disruptive event and the time after the event until recovery [39]. Enhancing resilience is measured by decreasing the area of the triangle formed by the system's performance function over time [40]. A variety of extensions to this basic concept have been proposed, including non-linear recovery [14], probabilistic assessments of resilience curves [41], and time-dependent resilience [42]. Community resilience could theoretically be measured in a similar manner to these metrics for infrastructure and engineering resilience [14], but measuring and even defining a community's performance before, during, and after a disruption is extremely challenging.

This article addresses several of these deficiencies with many of the current metrics and indices for community resilience. This article borrows from the existing literature on community resilience and infrastructure resilience to propose a different way to measure and assess community resilience through a VFT approach. VFT was developed specifically to help a decision maker select the best alternative for a multi-criteria decision problem [19]. VFT requires a decision maker to focus first on his or her values and objectives that he or she wants to achieve with a decision. VFT has been used to identify strategic objectives for an electric power utility [43], identify objectives and quantify the effectiveness of homeland security strategies [44], achieve consensus on decisions regarding the environment [43], and to assist communities in their planning [45, 46]. To the authors' knowledge, a VFT approach has not yet been applied to assessing community resilience.

Community resilience indices or measures are typically labeled as indicators because they are combinations of individual variables. Since the VFT approach also aggregates many variables, the approach presented in this article could also be called an indicator. We prefer the term metric to emphasize that we are seeking to measure community resilience. However, we do not intend to make a distinction between the use of the term indicator and metric.

## 3. A VALUE-FOCUSED THINKING APPROACH TO MEASURING COMMUNITY RESILIENCE

This article proposes to measure and assess community resilience by following VFT, which encourages decision makers to identify their fundamental objectives and attributes that define those fundamental objectives [47]. The first activity of VFT, identifying and structuring objectives, frequently results in an objectives hierarchy. An objectives hierarchy begins with a single objective, which in this case is to maximize community resilience. That objective is decomposed into multiple fundamental objectives. Each of these fundamental objectives is further decomposed into sub-objectives, and this process continues until the bottom level of the objectives hierarchy consists of measurable attributes or metrics [44]. VFT with identification of multiple objectives, formulation of value functions and assignment of weights has been effectively applied and demonstrated in the context of energy and homeland security [47-49]. For VFT to be effective, each attribute at the bottom level of the objectives hierarchy must have a measurable quantity associated with it. Using an objectives hierarchy fulfills two purposes: (i) eliminates the vagueness of community resilience, and (ii) provides information about the policy makers' real objectives (i.e., what they really are concerned about in the context of a disaster).

Community resilience generally relates to a community's ability to withstand, adapt to, and recover from disruptions. Our approach identifies six fundamental objectives for community resilience: (i) maximize social resilience, (ii) maximize economic resilience, (iii) maximize infrastructure resilience, (iv) maximize environmental resilience, (v) maximize availability and use of resources, and (vi) maximize post-disaster functionality of critical services. These six fundamental objectives are similar to the broad categories or capacities that many researchers have proposed to measure community resilience. The unique element about this article is that the attributes that are used to measure each of these six fundamental objectives are comprised of outcomes rather than inputs or characteristics of the community.

### 3.1 Social resilience

Social resilience is defined by a disruption's impacts on community residents. A community exists to benefit its residents, and community leaders want to protect and make sure those residents are resilient to disruptive events. Every measure of community or disaster resilience that we know of incudes metrics related to residents of a community. As depicted in Table 1, social resilience in this article is decomposed into three components: (i) socially vulnerable (SV) residents, (ii) non-SV residents, and (iii) psychological resilience. Each of these three components are further broken down into metrics or measurable attributes. Although the list of attributes is designed to capture the most important elements of social resilience, community leaders may identify other elements or metrics of social resilience about which they are concerned.

**Table 1- Attributes to measure social resilience**

| |
|---|
| 1. Social resilience |
|     1.1 Socially vulnerable (SV) resident |
|         1.1.1 Minimize fatalities |
|         1.1.2 Minimize injuries |
|         1.1.3 Minimize number of people requiring evacuation |
|         1.1.4 Minimize number of people requiring temporary, non-congregate housing |
|         1.1.5 Minimize number of people requiring long-term housing |
|     1.2 Non-SV residents |
|         1.2.1 Minimize fatalities |
|         1.2.2 Minimize injuries |
|         1.1.3 Minimize number of people requiring evacuation |
|         1.1.4 Minimize number of people requiring temporary, non-congregate housing |
|         1.1.5 Minimize number of people requiring long-term housing |
|     1.3 Psychological resilience |
|         1.3.1 Minimize residents' fear |
|         1.3.2 Minimize symptoms of post-traumatic stress disorder |
|         1.3.3 Minimize personal disruption of lifestyle |
|         1.3.4 Minimize inconvenience to residents |
|     1.4 Number of people relocating from the community |

A community's SV residents (e.g., lower income groups, racial minorities, the elderly) often suffer disproportionally from disruptions [50]. FEMA's standardized impacts focus on people with access and functional needs [121]. In light of this information, we separate the social resilience of SV residents from the social resilience of non-SV residents. This distinction allows community leaders to focus on the most vulnerable residents and those people who are most likely to be harmed by a disruption while also tracking metrics corresponding to the majority, or the non-SV, proportion of the community.

The attributes for both SV and non-SV residents of a community consist of fatalities, injuries, and number of people requiring evacuation, temporary housing, and long-term housing. All of these attributes are listed as standardized impacts in FEMA's National Threat and Hazard Identification and Risk Assessment (THIRA) [121]. Jonkman et al. [51] provide a model to estimate fatalities from small-probability, large-consequence events. This model could be used by communities to forecast the number of fatalities that may occur from a disruptive event.

The components for SV residents and for non-SV events each include two attributes for housing: (i) minimizing the number of residents displaced from their homes and (ii) maximizing the number of displaced residents who find new housing. The ideal for a community during a disruption would be that no residents are displaced from their houses. However, since that ideal is often unattainable, communities will desire that those displaced residents have adequate housing and shelter. Frameworks and models have been proposed to integrate different scientific perspectives into post-disaster decision making and housing recovery for SV residents [52].

Large-scale disruptions can have lasting effects on the social cohesion of a community [53], and the psychology of residents plays a very important role in averting further losses and in recovery efforts. Residents' sentiments toward their community provide predictive power in determining psychological resilience to hazards such as toxic waste, salinity, and volcanoes. An individual's socioeconomic disadvantage, which could be driven by race, unemployment, or economic status, are associated with a greater likelihood of psychiatric disorder. Thus, the third component under social resilience is the psychological resilience of community residents.

Psychological resilience will probably be the most important for intentional incidents such as a terrorist attack or a mass shooting, but it may also be important for recovery from natural disasters. From the psychological literature, adult resilience can be defined as the ability of adults who are exposed to a traumatic or highly disruptive event to maintain a relatively healthy functioning of their psychological, emotional, and physical states [54]. Our framework proposes four attributes to measure psychological resilience: fear, posttraumatic stress disorder, disruption to lifestyle, and inconvenience, which are similar to [55]. Surveying residents could be used to measure these attributes.

**3.2 Economic resilience**

The economy of a community is vital to its survival. In order to ensure the survival and good health of a community, it is important to protect the economy from the adverse consequences of the disaster. Increasing the economy's resilience helps protect the economy from damage and enables the economy to recover more quickly. Rose [56] quantifies economic resilience according to static resilience and dynamic resilience.

Economic losses are frequently divided into direct losses (including the cost of damaged and destroyed buildings and the loss of industrial functions) and the indirect losses (second and third-order effects that are induced by the direct losses). The hierarchy for economic resilience (Table 2) includes a metric in order to address direct losses. Direct losses typically include the cost of infrastructure damage, debris removal and reconstruction. However, indirect losses include cascading losses as a result of direct losses, infrastructure damage and loss of functionality which in turn lead to business, workforce, and income losses.

**Table 2- Attributes to measure economic resilience**

| |
|---|
| 2. Economic resilience |
|     2.1 Minimum direct losses ($) |
|     2.2 Business resilience |
|         2.2.1 Minimize number of business closures |
|         2.2.2 Minimize length of time of business closure |
|         2.2.3 Minimize number of businesses that cannot reopen |
|     2.3 Workforce resilience |

|   |   |
|---|---|
|   | 2.3.1 Minimize number of residents who cannot find jobs or work again |
|   | 2.3.2 Minimize time that residents cannot find work |
|   | 2.3.3 Minimize number of available jobs that cannot find suitable employees |
|   | 2.3.4 Minimize time until available jobs are filled |
| 2.4 Income losses |   |
|   | 2.4.1 Minimize income losses of SV residents |
|   | 2.4.2 Minimize income losses of non-SV residents |
|   | 2.4.3 Minimize residential losses that are not insured |

Many models used to measure losses, such as the input-output model, the social accounting matrix, and the computable general equilibrium model have evolved to incorporate disaster-specific factors [57]. Many studies have tailored input-output models to suit specific scenarios [58-60]. Hallegate [61] attempted to estimate the impacts of a natural disaster on the supply chain using an input-output inventory model. There are also methods to estimate the impact of a disaster induced supply chain constraint where input-output models are not applicable [62]. Martinelli et al. [63] provide a framework based on HAZUS to assess the economic impact of natural disasters.

Since the health of an economy is dependent on the financial health of both enterprises and residents, business resilience and income losses attempt to capture the effects of the event on businesses and residents, respectively. Business resilience focuses on the post-disaster operations or closures of businesses and on how the disaster has affected the workforce availability.

Surveys of business after a disaster can provide insight into how business are impacted by the event and on their speed and effectiveness of recovery. Zobel [64] applies the infrastructure resilience of Bruneau et al. [65] to quantify the resilience of businesses and organizations. He measures the resilience of business based on lost performance and the time to recover to full performance. Similarly, the number of business that close and the length of time that it takes them to reopen are the metrics in this article used to measure business resilience. The number of permanently closed businesses are also addressed in the hierarchy.

Most of the studies cited in the previous paragraphs focus on the economic impact of a disruption, but significant research has also modeled economic recovery. Webb et al. [66] argue that long-term recovery of businesses is affected by various firm characteristics, including the prevailing market conditions, physical damage and disruption of operations. Porter [67] argues that regional economies impacted by the *Deepwater Horizon* oil spill would recover more slowly because of the global economic recession. Stock markets have been shown to be resilient to shocks caused by earthquakes [68].

Income losses address the financial effects of the disaster on community residents. It considers many different sources of income like wages and rent. Some disasters may have minimal impact on total employment, but there can be significant drops in personal income [69]. This category has been further divided into income losses for SV and non-SV residents to focus attention that SV residents may suffer more from income losses than non-SV residents. Workforce resilience is measured as the number of people that are out of work and the time it takes for people to resume working. Workforce resilience also includes the loss of employees and the time it takes for businesses to find new employees.

### 3.3 Infrastructure resilience

The failure of infrastructure systems can cripple a community. Infrastructure systems may be very vulnerable to damage during natural disasters. We address the resilience of infrastructure systems in terms of the damage sustained by the systems i.e. impact and the time and effort required for their restoration i.e. recovery. As depicted in Table 3, infrastructure resilience is comprised of debris management and critical infrastructure resilience. These three categories of resilience will be further decomposed into their respective components.

The resilience of infrastructure systems is a popular topic for researchers in engineering. Some studies have proposed resilience indices specifically for infrastructure systems. Fischer et al.'s [70] resilience index (RI), ranging from 0 to 100, is derived from three categories: robustness, resourcefulness, and recovery. The inoperability input-output model, network models, and "fragility" functions have all been proposed to assess resilience [71,72]. Many frameworks assess the resilience of infrastructure systems in terms of two dimensions: robustness (ability to withstand impact) and rapidity (time to recovery) [73].

Post-disaster debris can cause further accidents and damage and will generally be an obstacle to recovery efforts [74]. Removing debris is necessary to facilitate the recovery of the affected region. This component is measured as the time taken to clear debris. Debris can be estimated based on the type of debris (e.g., structural, trees, sediment, mixed), the location, and volume of the structure [75]. Simulation can be used to model the time to remove debris based on the expected volume and the ability of debris management services.

The criticality of infrastructure components depends on which sets are damaged or destroyed by a disruption [76]. Table 3 depicts four critical infrastructures: transportation, energy, communication systems, and waste management, all of which are standardized impacts in the THIRA. These seem to be the most critical because of the amount of attention and focus that disaster researchers have spent studying these systems. Water infrastructure could be another critical infrastructure system, but that is included in the hierarchy for resource resilience. Policymakers can include more critical infrastructure systems in their measures for community resilience if they believe those systems are important and reflect their values.

Transportation resilience addresses three major modes of transportation: roadways, airways, and waterways. These transportation systems are very important for recovery efforts since they can be used for evacuations as well as to bring in additional resources. The impacts and recovery of these systems rely on metrics that are appropriate for the mode of transport. For example, the impact on roadways is measured by the number of road closures, and the impact of airways is measured by the number of cancelled flights.

A common approach to enhance resilience of transportation networks is the usage of stochastic modeling or stochastic programming or their variants [77-80]. Many studies measure the impacts on supply chain caused by the physical damage to transportation systems [81, 82]. Transportation disruptions can also lead to significant business interruption losses [83]. Chang and Nojima evaluate the post-disaster performance of Kobe city's transportation network in terms of network coverage and transport accessibility [84]. These types of studies can be used to understand how a community's transportation system might be impacted by different types of disruptive events.

To account for the different forms of energy available to a community resident, energy resilience is decomposed into electricity, gas, and fuel. The attributes consist of the impacts and recovery time for both residential customers and business customers. Research in energy resilience approximate the impact of disruptions on energy supply. MacKenzie and Barker [85] provide a data-driven approach to derive a resilience parameter through regression models with electric power outage data. Spatial generalized linear mixed modeling applied to grid cells in a region can be used to predict the number of outages likely to occur as a result of storms [86]. Damage to the electric power system due to hurricanes can be assessed by modeling the expected damage to electric poles [87].

**Table 3- Attributes to measure infrastructure resilience**

| |
|---|
| 3. Infrastructural resilience |
| 3.1 Minimize time to clear debris and remove damaged buildings and infrastructure |
| 3.2 Critical infrastructure resilience |
|     3.2.1 Transportation resilience |
|         3.2.1.1 Highway and road resilience |
|             3.2.1.1.1 Minimize miles of highway and road closure |
|             3.2.1.1.2 Minimize time that highways and roads are closed |
|         3.2.1.2 Airport resilience |
|             3.2.1.2.1 Minimize number of cancelled flights |
|             3.2.1.2.2 Minimize time to recover normal airport operations |
|         3.2.1.3 Waterway resilience |
|             3.2.1.3.1 Minimize number or percentage of waterway port closures |
|             3.2.1.3.2 Minimize time until ports reopen or return to full operations |
|     3.2.2 Energy resilience |
|         3.2.2.1 Minimize number of customers without electricity |
|         3.2.2.2 Minimize time that customers do not have electricity |
|         3.2.2.3 Minimize number of customers without gas |
|         3.2.2.4 Minimize time that customers do not have gas |
|         3.2.2.5 Maximize availability of fuel (i.e., gasoline) |
|     3.2.3 Communications and information technology resilience |
|         3.2.3.1 Minimize number of telephone lines or poles damaged |
|         3.2.3.2 Minimize time to repair telephone lines or poles |
|         3.2.3.3 Minimize number of people who lose internet connectivity |
|         3.2.3.4 Minimize time to restore internet connectivity |
|     3.2.4 Waste management resilience |
|         3.2.4.1 Minimize sewage line closures |
|         3.2.4.2 Minimize time to restore sewage line closures |

Communication, in some form, is an important component of most if not all community resilience models [88, 89]. Communication systems and resources represent the reservoirs in which community meaning-making, information exchange, interactions, and connections can occur [90]. Communication technologies are extremely important in mitigating and preventing disasters [91]. Communication and information systems are also important for coordination operations during and after the disaster. The objectives hierarchy focuses on telephone poles and Internet connectivity. Damage to communication networks can be assessed through field collected data and information and by garnering availability [92]. In the hierarchy, communication resilience seeks to minimize impacts and minimize recovery time for Internet systems and telephone poles.

The resilience of waste management capabilities is measured by the impact and recovery of sewage lines in the affected region. Basic sanitation facilities and access to basic hygiene may be unavailable or worsen due to natural disasters [93]. Waste that is not properly managed are a serious health hazard and can further the spread of infectious diseases [94].

**3.4 Environmental resilience**

Disruptive events can also damage the environment. The specific attributes that measure the environment are likely very geographic-specific. However, some simple attributes that are applicable to a wide range of communities and locations are the geographic area of natural habitat, the number of animals impacted, and pollution.

Damage to the environment can often lead to the extinction or exodus of different plants and animals that could be crucial to the local ecosystem. Significant biomass decline facilitated by tree mortality and tree injury is one of the immediate effects of an earthquake [95]. The 2004 Indian ocean tsunami led to changes and uprooting in the mangrove population due to seawater inundation [96]. The resilience framework presented in Table 4 assesses environmental impacts as the acreage of the habitat destroyed and the time for the habitat to recover. The recovery of habitats may be very different compared to other recoveries. Habitats sometimes take decades to recover to a pre-disaster state. The habitat may never be restored to its pre-disruption state. In such cases, recovery can be measured as the time until the community adapts to the "new" habitat. Table 4 also includes the impacts on individual animals. These animals can be further categorized into species depending upon their importance in maintaining the ecosystem balance.

**Table 4- Attributes to measure environmental resilience**

| |
|---|
| 4. Environmental resilience |
|     4.1 Minimize square miles of habitat destroyed |
|     4.2 Minimize time until habitat is restored |
|       4.3 Minimize number of animals impacted (could be categorized according to species) |
|       4.4 Pollution |
|           4.4.1 Minimize pollution in air |
|           4.4.2 Minimize pollution in water |
|           4.4.3 Minimize pollution in soil |

The impacts on the environment such as pollution should be measured in appropriate units such as parts per million for air and water pollution. Forest fires and fires from earthquakes and volcanic eruptions pollute the air and water. Volcanic eruptions are notorious for emitting vast quantities of polluting gases and ash resulting in global temperature changes [97, 98]. Floods can contaminate the soil and even saturate it with water. Given the importance of soil fertility and stability to agriculture and to construction projects, minimizing soil pollution is included as one of the objectives in environmental resilience.

## 3.5 Resource resilience

Resources include consumables like food and water and the sources of these consumables such as agriculture and livestock. Resource resilience (Table 5) addresses agriculture, food, and potable water. Agriculture resilience is further decomposed into metrics that measure the yield lost as a result of the disruption and the time taken to restore the pre-disruption state of agricultural yield. Agriculture resilience also aims to minimize the loss of livestock. If the disruption impacts a rural area where agriculture is a major component of the community, the impact and time to recovery of agriculture may be very important to assess community resilience. If the region is urban, the decision maker may not place much importance on agricultural resilience and can choose to focus more on food resilience. Resource shortages also lead to broader economic consequences over a period of time [99].

Resource resilience aims to minimize food shortages. Food resilience is measured in terms of amount of food shortage and the time it takes to end the shortage. Israel and Briones' [100] study in the Philippines found that typhoons negatively impact paddy rice production and the food security of the households in the affected areas. Tropical cyclones, floods, and droughts can also substantially impact natural resources. Natural disasters can affect multiple dimensions of food security such as the availability of supplies, access to food, and utilization. People in remote areas often suffer disproportionally from significant shortfalls in food availability [101].

**Table 5- Attributes to measure resource resilience**

| |
|---|
| 5. Resource resilience |
|     5.1 Agricultural resilience |
|         5.1.1 Minimize agricultural yield loss |
|         5.1.2 Minimize time to recover agricultural loss (e.g., harvest cycles) |
|         5.1.3 Minimize loss of livestock |
|     5.2 Food resilience |
|         5.2.1 Minimize number of people without sufficient food |
|         5.2.2 Minimize time until food shortage ends |
|     5.3 Maximize availability of portable water |

Given the importance of potable water to sustenance of human life and activity, resource resilience also seeks to maximize the availability of potable water to community residents. Aubuchon and Morley [102] assess the monetary benefit of continuing to provide water after a disruptive event to both businesses and residents. Luna et al. [103] use colored Petri nets to simulate the behavior and restoration process of a water distribution network in Tokyo following an earthquake. Water supply networks may also be vulnerable to physical attacks [104].

**3.6 Post-disaster functionality of critical services**

Emergency services are critically important to mitigate the effects of a disruption. However, the services themselves can also be susceptible to the effects of a disruption either because the disruption directly affects these services or because the services' capabilities are overstretched by the disruption. The objectives hierarchy (Table 6) includes the ability of the medical, police, fire, educational, and social services to continue to provide necessary functions during and after a disaster.

The functionality of medical services can be assessed as a percentage of the pre-disaster condition. Medical services must consider the availability of both personnel and emergency medical equipment for considerably high emergency patient traffic. Hospital emergency departments throughout the United States are severely crowded, which raises concerns about their ability to respond to mass casualty or volume surges [105]. Many medical facilities that would need to respond to a disaster might have inadequate disaster plans [106, 107]. Medical facilities can also be damaged by the disruptive event, which can lead to the loss of vital services, as occurred during the 1994 Northridge earthquake [108].

The functionality of the police and fire departments can be measured based on the number of available personnel after the disaster. The personnel and equipment of these services can be physically impacted by the disruption, reducing their effectiveness, and potentially rendering them ineffective. Emergency workers pressed into service during times of crisis are seriously affected by the emergency work [109]. According to surveys, emergency personnel may not participate equally in the response to different threats [110].

**Table 6- Attributes to measure functionality of critical services**

| |
|---|
| 6. Post-disaster functionality of critical services |
|     6.1 Medical services |
| 6.1.1 Maximize ratio of post-disruption capability to pre-disruption capability |
| 6.2 Police services |
| 6.2.1 Maximize number of law enforcement officers available post-disruption |
| 6.3 Fire management services |
| 6.3.1 Maximize number of firefighters available post-disruption |
| 6.4 Education services |
| 6.4.1 Maximize number of schools open post-disaster |
| 6.4.2 Minimize amount of time until all schools are reopened |
| 6.4.3 Maximize number of students who attend schools |
| 6.5 Social, safety-net services |
| 6.5.1 Maximize number of employees working in social, safety-net services |

Education plays a central role in a community. Apart from being centers of learning, they routinely serve as designated shelters during a disruption [111]. Disruptions can negatively impact the educational function of schools because they mentally affect students and cause disturbances in coursework [112]. It is the interest of students, their families, and the community to restore schools' functioning as soon as possible.

For residents who are unable to work or unable to support themselves, a natural disaster makes matters much worse. Social safety nets are an absolute necessity for such residents. Another metric of resilience is the functioning of safety net services as they offer social protection and social risk management, thereby reducing impact and aiding in recovery [113].

## 4. RESILIENCE FRAMEWORK FOR DECISION MAKING

### 4.1 Operationalizing objectives for community resilience

Perhaps the biggest challenge with using this objectives hierarchy and these attributes to assess community resilience is the difficulty in assigning a number for each attribute, especially prior to a disruption. Since a resilience index and associated metrics should inform decision making and risk management, state and local government officials need to be able to assess a community's resilience before a disruption and understand how resilience can be enhanced through emergency preparedness.

Modeling and analytical tools can help a community assign numbers to these attributes. Data from previous disruptions—both disruptions experienced by the community and disruptions in other locations—could be used to assess each attribute. Simulation provides a powerful method to understand the impacts of different types of disruptions with varying degrees of severity. As cited in the Section 3, mathematical models have been proposed to describe how disruptions impact the performance of specific systems (e.g., transportation infrastructure, economic activity). Assigning probability distributions can quantify the uncertainty that usually exists in each of these attributes.

Until now, the discussion has centered around the first step of VFT, identifying and structuring objectives. We defined objectives and the metrics that measure progress towards the objectives. In order to assess a community's resilience, VFT recommends combining all of the attributes into a single number through a multi-attribute value function or utility function. The specifics of the aggregation scheme including the value or utility functions and the trade-off weights for each attribute will depend on the specific community. Decision analysis has provided numerous examples demonstrating how to derive value and utility functions and trade-off weights for a wide variety or private-sector and public-sector decision problems in which the problems contain scores of attributes. These same techniques can also be applied to help a community leader construct a resilience metric that aggregates the attributes in the objectives hierarchy.

**4.2 Illustrative numerical example**

An illustrative numerical example serves to demonstrate the application of the proposed community resilience framework to evaluate a community's resilience to diverse disruptive events.

A Midwest city with approximately 250,000 residents is considering five disruptive events as part of its emergency planning process. The events are comprised of two intentional attacks—an IED attack in a downtown area and a cyberattack—and three weather events—a tornado, flood, and winter storm. We assume the city has extensive experience dealing with the weather events but has never experienced an IED attack or a cyberattack that affects a large proportion of the city. The city serves as the state's capital, and a large percentage of its residents are professionals in the administrative, healthcare, education, and financial industries. Its median household income is $60,000, but 15% of its residents live below the poverty line.

*4.2.1 Multi-attribute value function for resilience*

We assume the city leaders use the objectives hierarchy attributes described in Section 3 but do not include the two waterway resilience attributes (infrastructure resilience) or the three agricultural resilience attributes (resource resilience) because these attributes do apply to their city. Thus, there are 56 attributes that will comprise the city's resilience assessment. Many attributes in the objectives hierarchy are mutually preferential independent. Attributes are mutually preferential independent if an individual value function does not depend on the specific levels or trade-offs among the other attributes. For attributes that are mutually preferential independent, a linear or exponential value function can reflect the city officials' preferences if they believe the attribute provides constant returns to scale (linear) or increasing or decreasing returns to scale (exponential).

Since several components of resilience—especially within infrastructure resilience—consist of an attribute describing the impact and another attribute describing recovery time, the attributes of impact and recovery time are not mutually preferentially independent. The resilience triangle, developed by Bruneau et al. [39] provides a method to use the product of the magnitude of impact and recovery time. Zobel [36] measures resilience as the area under the curve (i.e., the triangle) which is normalized by the maximum impact and the maximum time in order to ensure resilience is bounded between 0 and 1. Based on the concept of the resilience triangle, a value function based on the product of impact $x_{impact}$ and recovery time $x_{time}$ can be used:

$$v_{impact,time}(x_{impact}, x_{time}) = 1 - \frac{x_{impact} * x_{time}}{\bar{x}_{impact} * \bar{x}_{time}} \quad (1)$$

where $\bar{x}_{impact}$ is the maximum tolerable impact and $\bar{x}_{time}$ is the maximum tolerable recovery time. The city leadership's value functions (linear, exponential, or Equation 1) appear in Tables S1-S6 in the supplementary information. Every value function $v(\cdot)$ is bounded between 0 and 1. If the function evaluates to less than 0, $v(\cdot) = 0$, and if the value function evaluates to more than 1, $v(\cdot) = 1$.

Since 24 attributes are paired together via Equation 1 and the other 32 attributes are mutually preferential independent, we have 44 individual value functions. Given the set of attributes **x** that make up community resilience, the city's resilience $R$ can be calculated:

$$R(\mathbf{x}) = \sum_{\forall i} w_i v_i(x_i) \quad (2)$$

where $0 \leq w_i \leq 1$ is the trade-off weight corresponding to attribute $i$ (or a single pair of attributes), and $0 \leq v_i(x_i) \leq 1$ is the value function evaluated at level $x_i$ for attribute $i = 1, 2, \ldots, 44$.

Weights for each attribute or for each combined value of impact and recovery time are derived for each section and sub-section of the objectives hierarchy. We assume the city leaders determine their weights using different methods including swing weighting and rank reciprocal. Table 7 depicts the weights assigned to the six high-level objectives. City leaders determine that social resilience, resource resilience, and post-disaster functionality of critical service are most important for enhancing their city's community resilience.

| Table 7- Weights for each objective ||
| Objective | Weight |
| --- | --- |
| Social resilience | 0.289 |
| Economic resilience | 0.159 |
| Infrastructure resilience | 0.191 |
| Environmental resilience | 0.051 |
| Resource resilience | 0.180 |
| Functionality of critical services | 0.130 |

*4.2.2 Results*

We model the uncertainty in our evaluation of each of the 56 metrics for resilience with probability distributions. Each metric is assessed with either a normal or a lognormal distribution depending on whether we want the uncertainty to be symmetric or right-skewed. The state of Iowa's THIRA provides numerical estimates of standardized impacts for each of the five disruptive events, and several of the resilience metrics are the same as the THIRA standardized impacts. Since the THIRA standardized impacts represent worst-case scenarios, we assume the standardized impact is the 95th percentile for a metric. For example, if the THIRA estimates 100 fatalities for a disruptive event, we assume $P(X \leq 100) = 0.95$ where $X$ is the number of fatalities from the disruptive event. Since a disruptive event may not lead to disastrous outcomes (e.g., nobody dies from a tornado), we choose mean values for these distributions so that there is a significant probability of relatively minor outcomes. The authors' best estimates for parameters are assigned to metrics that do not align with a standardized impact. Since we expect the outcomes to be highly correlated, we assume a 0.7 correlation coefficient between each attribute.

Monte Carlo simulation is used to randomly generate 1000 instances for each metric for each disruptive event. The city's resilience is calculated for each instance based on the 56 metrics, and each disruptive event results in 1000 resilience assessments between 0 and 1. The results of these simulated resilience assessments are presented in Figure 1 as a box plot to depict the distributions and variability of resilience across the different events. The white diamond in the middle of the box represents the average resilience value, and the circles extending beyond the lines are outliers in the simulation. Table 8 depicts the minimum, maximum, and average values for each disruptive event.

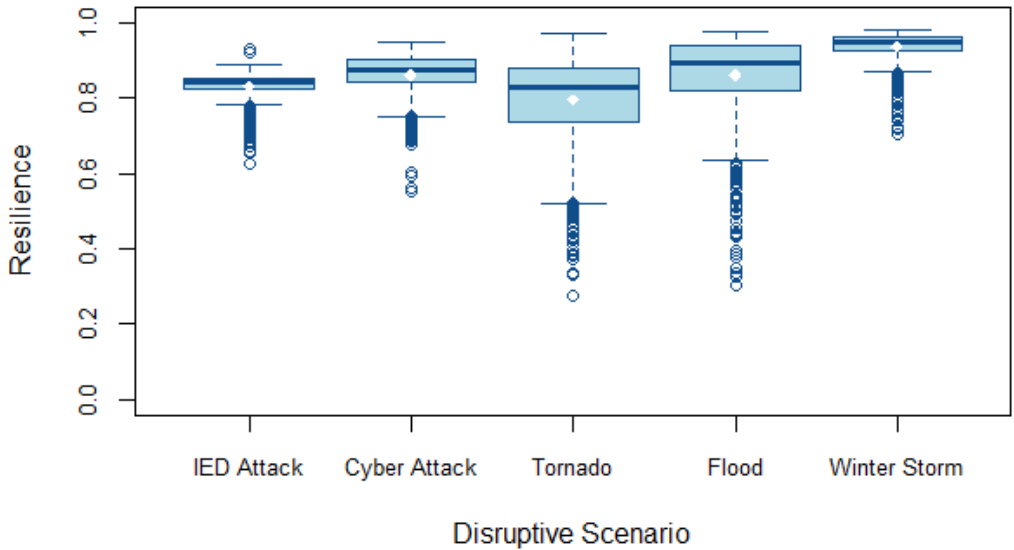

**Figure 1- Boxplot of simulated community resilience for five disruptions**

**Table 8- Summary statistics for community resilience**

| Disruptive Event | Resilience Range | Average |
|---|---|---|
| IED attack | (0.63, 0.93) | 0.83 |
| Cyberattack | (0.55, 0.95) | 0.86 |
| Tornado | (0.28, 0.97) | 0.79 |
| Flood | (0.30, 0.97) | 0.86 |
| Winter storm | (0.70, 0.98) | 0.94 |

The results show that the city's average resilience is between 0.79 and 0.94 for the five disruptive events. On average, the city is most resilient to a winter storm and least resilient to a tornado. Community resilience is most variable for tornado and flood events. Frequently, the impacts from a tornado or flood are relatively minor but there are some instances when these events can generate substantial destruction and fatalities. The city is most resilient to a winter storm, indicating a high preparedness and adaptability for the city's infrastructure and resources when confronted with winter-storm hazards.

Comparing the resilience of the IED attack and a flood generates important insights. The city's average resilience to a flood is greater than its average resilience to an IED attack. The city is experienced in dealing with floods, and most flooding will be relatively minor. The city's lack of preparedness for an IED attack means that on average the city is less resilient to an IED attack than to a flood. The minimum resilience for an IED attack is 0.63, but the flooding simulation results in several scenarios in which community resilience is less than 0.6. A large IED attack on the scale of the 1995 Oklahoma City bombing would still only cause damage in a relatively small geographic area, but a major flood could damage the entire city.

The city's resilience to each disruptive event can be disaggregated according to the six criteria in order to understand the contribution of each objective to the overall resilience and identify areas where the city may be more vulnerable to a disruptive event. Figure 2 depicts the percentage to the ideal resilience value for each objective for each disruptive event, based on the average resilience values. As shown in Table 2, the city is least resilient to a tornado. Figure 2 shows that the city performs relatively badly on social, economic, and environmental resilience and functionality of critical services for a tornado, with values less than 80% of the ideal resilience for those four criteria. The city is more resilient to a flood but it only achieves 60% of the ideal values for environmental resilience. The city's infrastructure is most vulnerable to a cyber attack, and its infrastructure resilience is 54% of its ideal infrastructure resilience. The city's social resilience for an IED attack is 71% of its ideal value. City emergency preparedness officials can use the analysis to determine which areas might need more resources to increase their resilience. A tornado requires resources across all several areas. An IED attack requires more resources for social resilience (e.g., human lives), and a cyber attack requires more resources for its infrastructure.

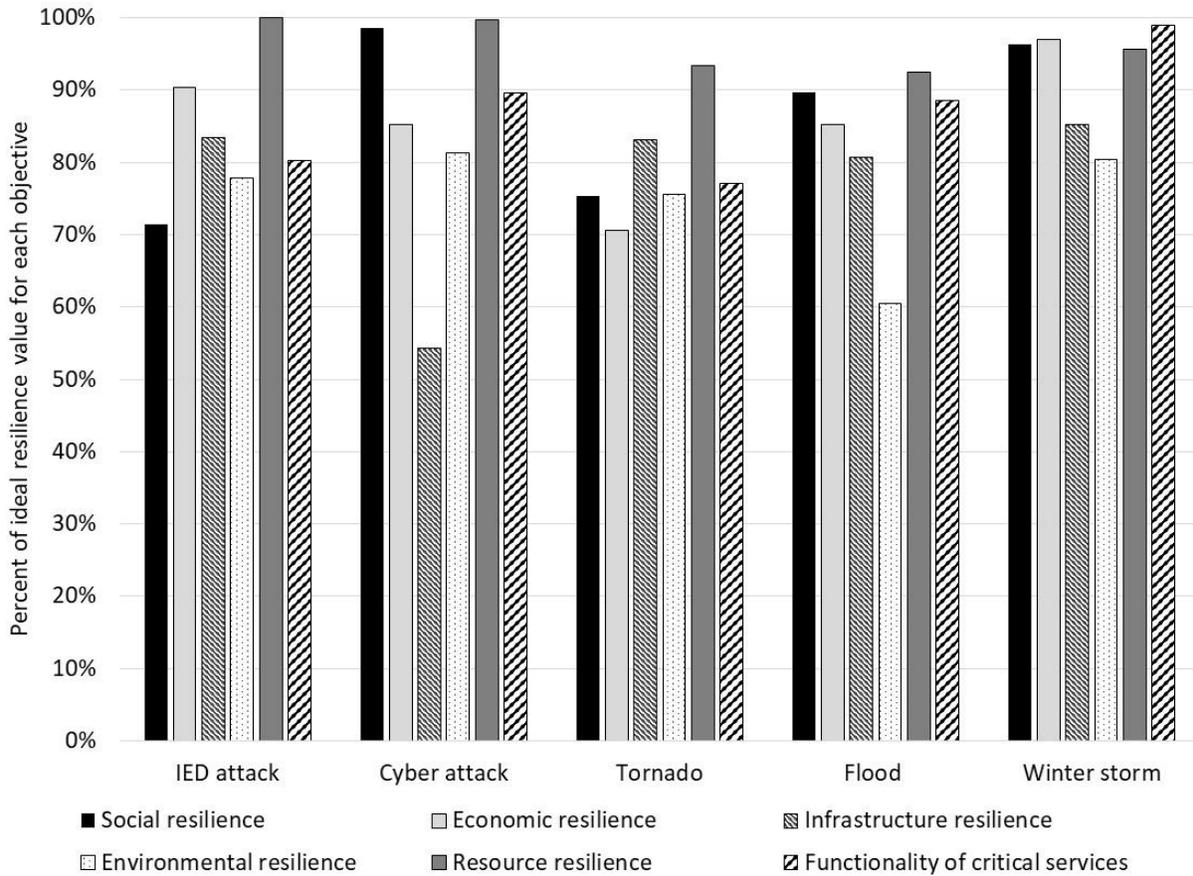

**Figure 2- Resilience values for each objective as percent of ideal**

## 5. CONCLUSION

This article aims to apply VFT to the domain of disaster resilience. It employs an objectives hierarchy to clearly define the objectives of a decision maker. The objectives hierarchy focuses on fundamental objectives of protecting and recovering parts of a community that are necessary for proper functioning: people, economy, infrastructure, environment, resources, and emergency services. The combination of these objectives provide insight into the resilience of a community.

Each of these six objectives are decomposed into sub-objectives and result in measurable attributes. Unlike most measures for community resilience currently found in the literature, the attributes proposed in this work consist of outcomes from a disruptive event as opposed to inputs or characteristics of a community.

This paper includes scores of attributes that communities may want to consider in assessing resilience. A VFT approach to community resilience means that this framework can be tailored for each community. Communities will have different value functions and importance weights that reflect community characteristics and their priorities. Coastal communities may emphasize waterway resilience, and rural communities may emphasize agriculture resilience. Large metropolitan cities with diverse population groups may emphasize social resilience or include more attributes that explicitly measure the impacts of a disruption on diverse population groups.

As the numerical example illustrates, a VFT approach to community resilience enables a community to assess its resilience to individual disruptive events based on numerically assessing different outcomes from potential disruptions. Since uncertainty will exist with these metrics, Monte Carlo simulation can be deployed to calculate resilience that incorporates the uncertainties inherent in each attribute and hazard. The example shows that a community's average resilience may be greater for one hazard (e.g., flood) than for another hazard (e.g., IED attack), but the hazard whose average resilience is greater could have more variation than other hazard. The former hazard may result in scenarios that generate significant disruptions that cripple a community even if the typical or average scenario is relatively mild. Other resilience metrics based on census and geographic data do not generate this type distinction between hazards or within a disruptive event.

Humans have developed technologies to overcome many problems, but natural disasters continue to be a challenge. Since preventing natural events is usually impossible, we should focus our efforts towards making our communities more resilient towards disruptions. Many researchers have worked on measuring and improving community resilience, but community leaders still struggle with determining how to implement strategies to enhance their communities' resilience. This article constructs measures focused on what stakeholders' value in order to provide more meaningful and more actionable measures for resilience.

# SUPPLEMENTARY INFORMATION

The supplementary information provides more details about the value functions and weights that are used to measure resilience for the illustrative numerical example in Section 4.2. The illustrative example makes use of three types of value functions: linear, exponential, and mutually dependent. Each value function is bounded between 0 and 1 and can be tailored for attributes in which more of an attribute is better and for attributes in which less of an attribute is better. A linear value function $v_i(x_i)$ in which more of attribute $x_i$ is better is:

$$v_i(x_i) = \begin{cases} 0 & \text{if } x_i \leq x^- \\ \dfrac{x_i - x^-}{x^+ - x^-} & \text{if } x^- < x_i \leq x^+ \\ 1 & \text{otherwise} \end{cases}$$

where $x^-$ is the worst case of an attribute and $x^+$ is the best case for an attribute. A linear value function in which less of an attribute $x_i$ is better is:

$$v_i(x_i) = \begin{cases} 1 & \text{if } x_i \leq x^+ \\ \dfrac{x^- - x_i}{x^- - x^+} & \text{if } x^+ < x_i \leq x^- \\ 0 & \text{otherwise} \end{cases}$$

An exponential value function in which more of attribute $x_i$ is better is:

$$v_i(x_i) = \begin{cases} 0 & \text{if } x_i \leq x^- \\ \dfrac{1 - \exp(\alpha[x_i - x^-])}{1 - \exp(\alpha[x^+ - x^-])} & \text{if } x^- < x_i \leq x^+ \\ 1 & \text{otherwise} \end{cases}$$

where $\alpha$ describes the curvature of the value function. An exponential value function in which less of an attribute $x_i$ is better is:

$$v_i(x_i) = \begin{cases} 0 & \text{if } x_i \leq x^+ \\ \dfrac{1 - \exp(\alpha[x^- - x_i])}{1 - \exp(\alpha[x^- - x^+])} & \text{if } x^+ < x_i \leq x^- \\ 1 & \text{otherwise} \end{cases}$$

The mutually dependent value function is provided in Equation 2 in the main body of the article.

Tables S1-S6 provide the units, the best case and worst case, and the type of value function used for each attribute. Each table provides the attributes for one of the six main objectives. Table S7 provides the trade-off weights for each attribute. The individual value functions and the weights combine to measure community resilience according to Equation 1.

Table S1- Value functions of social resilience attributes

| Attribute | Units | Best case | Worst case | Value function |
|---|---|---|---|---|
| Fatalities (SV residents) | Count | 0 | 50 | Linear |
| Injuries (SV) | Count | 0 | 1000 | Linear |
| People requiring evacuation (SV) | Count | 0 | 10,000 | Linear |
| People requiring temporary, non-congregate housing (SV) | Count | 0 | 500 | Linear |
| People requiring long-term housing (SV) | Count | 0 | 500 | Linear |
| Fatalities (non-SV residents) | Count | 0 | 500 | Linear |
| Injuries (non-SV) | Count | 0 | 10,000 | Linear |
| People requiring evacuation (non-SV) | Count | 0 | 50,000 | Linear |
| People requiring temporary, non-congregate housing (non-SV) | Count | 0 | 1000 | Linear |
| People requiring long-term housing (non-SV) | Count | 0 | 1000 | Linear |
| Residents' fear | % residents | 0 | 75 | Linear |
| Symptoms of PTSD | % residents | 0 | 30 | Linear |
| Personal disruption of lifestyle | % residents | 0 | 100 | Linear |
| Number of people who relocate from the community | % residents | 0 | 20 | Linear |

Table S2- Value functions of economic resilience attributes

| Attribute | Units | Best case | Worst case | Value function |
|---|---|---|---|---|
| Direct losses | Dollars | 0 | 1 million | Exponential |
| Number of business closures | Count | 0 | 400 | Mutually dependent |
| Length of time of business closure | Days | 0 | 365 | Mutually dependent |

| Attribute | Units | Best case | Worst case | Value function |
|---|---|---|---|---|
| Number of permanent business closures | Count | 0 | 100 | Linear |
| Number of residents who cannot find jobs or work again | % residents | 0 | 10 | Linear |
| Number of available jobs that cannot find suitable employees | % of jobs | 0 | 20 | Mutually dependent |
| Time until available jobs are filled | Days | 0 | 365 | |
| Income losses of SV residents | Dollars | 0 | 15 million | Linear |
| Income losses of non-SV residents | Dollars | 0 | 100 million | Linear |
| Residential losses that are not insured | Dollars | 0 | 200 million | Linear |

Table S3- Value functions of infrastructure resilience attributes

| Attribute | Units | Best case | Worst case | Value function |
|---|---|---|---|---|
| Time to clear debris | Days | 0 | 90 | Exponential |
| Miles of highway and road closures | Miles | 0 | 80 | Mutually dependent |
| Time for which highways and roads are closed | Days | 0 | 30 | |
| Number of daily cancelled flights | Count | 0 | 160 | Mutually dependent |
| Time to restore normal airport operations | Days | 0 | 90 | |
| Number of customers without electricity | Count | 0 | 100,000 | Mutually dependent t |
| Time that customers do not have electricity | Days | 0 | 21 | |
| Number of customers without gas | Count | 0 | 100,000 | Mutually dependent t |
| Time that customers do not have gas | Days | 0 | 21 | |
| Number of commercial buildings without gas | Count | 0??? | 5000??? | Mutually dependent |
| Time that commercial buildings do not have gas | Days | 0??? | 60??? | |
| Availability of fuel (gasoline) | % of operational gas stations | 100 | 80 | Linear |
| Number of telephone lines/poles damaged | Count | 0 | 10,000 | |

| Attribute | Units | Best case | Worst case | Value function |
|---|---|---|---|---|
| Time to repair telephone lines/poles | Days | 0 | 180 | Mutually dependent |
| Number of people who lose internet connectivity | Count | 0 | 100,000 | Mutually dependent |
| Time to restore internet connectivity | Days | 0 | 21 | |
| Number of sewage line closures | Count | 0 | 150 | Mutually dependent |
| Time to restore sewage lines | Days | 0 | 21 | |

**Table S4- Value functions of environmental resilience attributes**

| Attribute | Units | Best case | Worst case | Value function |
|---|---|---|---|---|
| Square miles of habitat destroyed | Sq. miles | 0 | 2 | Mutually dependent |
| Time until habitat is restored | Projected years | 0 | 15 | |
| Number of animals impacted | Count | 0 | 7500 | Linear |
| Pollution in air | Air quality index | 150 | 0 | Linear |
| Pollution in water | Water quality index | 90 | 25 | Linear |
| Pollution in soil | Soil quality index | 100 | 20 | Linear |

**Table S5- Value functions of resource resilience attributes**

| Attribute | Units | Best case | Worst case | Value function |
|---|---|---|---|---|
| Number of people without sufficient food | Count | 0 | 50,000 | Mutually dependent |
| Time until food shortage ends | Days | 0 | 30 | |
| Customers without water service | Count | 0 | 50,000 | Exponential |

**Table S6- Value functions of post-disaster functionality of critical services**

| Attribute | Units | Best case | Worst case | Value function |
|---|---|---|---|---|
| Ratio of post-disruption capability to pre-disruption capability of medical services | Ratio | 1 | 0.7 | Linear |
| Number of law-enforcement officers available post-disruption | % of pre-disaster availability | 100 | 80 | Linear |
| Number of firefighters available post-disruption | % of pre-disaster availability | 100 | 80 | Linear |
| Number of schools open post-disaster | % of pre-disaster availability | 100 | 70 | Mutually dependent |
| Time until all schools are reopened | Days | 0 | 90 | |
| Number of students who attend schools | % of pre-disaster enrollment | 100 | 70 | Linear |
| Number of employees working in social, safety-net services | % of pre-disaster availability | 100 | 70 | Linear |

**Table S7- Weight for each attribute**

| Attributes | Weights |
|---|---|
| Fatalities (SV residents) | 0.048 |
| Injuries (SV) | 0.019 |
| People requiring evacuation (SV) | 0.030 |
| People requiring temporary, non-congregate housing (SV) | 0.011 |
| People requiring long-term housing (SV) | 0.022 |
| Fatalities (non-SV residents) | 0.048 |
| Injuries (non-SV) | 0.019 |
| People requiring evacuation (non-SV) | 0.030 |
| People requiring temporary, non-congregate housing (non-SV) | 0.011 |
| People requiring long-term housing (non-SV) | 0.022 |
| Residents' fear | 0.006 |
| Symptoms of PTSD | 0.012 |
| Personal disruption of lifestyle | 0.002 |
| Number of people who relocate from the community | 0.009 |
| Direct losses | 0.08 |
| Number of business closures | |
| Length of time of business closure | 0.012 |
| Number of permanent business closures | 0.012 |
| Number of residents who cannot find jobs or work again | 0.016 |
| Number of available jobs that cannot find suitable employees | |
| Time until available jobs are filled | 0.008 |
| Income losses of SV residents | 0.011 |
| Income losses of non-SV residents | 0.010 |
| Residential losses that are not insured | 0.010 |
| Time to clear debris | 0.038 |
| Miles of highway and road closures | 0.022 |
| Time for which highways and roads are closed | |
| Number of daily cancelled flights | 0.020 |
| Time to restore normal airport operations | |
| Number of customers without electricity | 0.025 |
| Time that customers do not have electricity | |
| Number of customers without gas | 0.016 |
| Time that customers do not have gas | |
| Availability of fuel (gasoline) | 0.005 |
| Number of telephone lines/poles damaged | 0.021 |
| Time to repair telephone lines/poles | |
| Number of people who lose internet connectivity | 0.021 |
| Time to restore internet connectivity | |
| Number of sewage line closures | 0.023 |
| Time to restore sewage lines | |
| Square miles of habitat destroyed | 0.015 |
| Time until habitat is restored | |
| Number of animals impacted | 0.015 |
| Pollution in air | 0.009 |
| Pollution in water | 0.009 |
| Pollution in soil | 0.003 |
| Number of people without sufficient food | 0.081 |
| Time until food shortage ends | |
| Customers without water service | 0.099 |
| Ratio of post-disruption capability to pre-disruption capability of medical services | 0.026 |
| Number of law-enforcement officers available post-disruption | 0.026 |
| Number of firefighters available post-disruption | 0.026 |
| Number of schools open post-disaster | 0.015 |
| Time until all schools are reopened | |
| Number of students who attend schools | 0.011 |
| Number of employees working in social, safety-net services | 0.026 |